\documentclass[amsmath,floats,floatfix,nofootbib, notitlepage, preprint]{revtex4-1}
\DeclareMathAlphabet{\mathpzc}{OT1}{pzc}{m}{it}
\DeclareMathAlphabet{\mathscrligra}{T1}{calligra}{m}{n}
 \usepackage{hyperref}
\usepackage{graphicx}
\usepackage{amsmath}
\usepackage{amsfonts}
\usepackage{calligra}
\usepackage{mathrsfs}
\usepackage{bm}
\usepackage{tensind}
\usepackage{feyn}
\usepackage{PeterStyle}

\begin{document}
\title{Gravitational self-force in scalar-tensor gravity}
\author{Peter~Zimmerman\footnote{{\tt pzimmerm@uoguelph.ca}}}
\affiliation{Department of Physics, University of Guelph}
\date{\today}

\begin{abstract}
Motivated by the theoretical possibility of floating orbits and the potential to contribute extra constraints on alternative theories, in this paper we derive the self-force equation for a small compact object moving on an accelerated world line in a  background spacetime which is a solution of the 
coupled gravitational and scalar field equations of scalar-tensor theory.  In the Einstein frame, the coupled field equations governing the perturbations sourced by the particle share the same form as the field equations for perturbations of a scalarvac spacetime in general relativity, with both falling under the general class of hyperbolic field equations studied in \cite{Zimmerman:2014uja}. Here, we solve the field equations formally in terms of retarded Green functions, which have explicit representations as Hadamard forms in the neighbourhood of the world line.  Using a quasi-local expansion of the Hadamard form, we derive the regular solutions in Fermi normal coordinates according to the Detweiler-Whiting prescription. To compute the equation of motion, we parameterize the world line by the particle's mass and ``charge'', which we define in terms of the original Jordan frame mass, its derivative, and the parameter which translates the proper time in the Jordan frame to the Einstein frame.  These parameters  depend on the value of the background scalar field and its self-field corrections.  The equation of motion which follows from the regular fields  strongly resembles the equation for the self-force acting on a charged, massive particle in a scalarvac geometry of general relativity. Unlike the scalar vacuum scenario, the ``charge'' parameter in the scalar-tensor self-force equation is time variable and leading to  additional local and tail terms.  We also provide evolution equations for the world line parameters under the influence of the self-fields.
  
\end{abstract}
\maketitle 

\section{introduction}
\subsection{Motivation for alternative theories}
The general theory of relativity (GR) is widely considered to be the most successful classical theory of gravitational phenomena. Its predictions have passed many stringent tests such as the light deflection around the sun during a solar eclipse, the Shapiro time delay, the perihelion advance of Mercury, and the Nordtvedt effect in lunar motion \cite{Will:2001mx}.  Despite its great success in predicting deviations from Newtonian gravity in the solar system where deviations from flat spacetime are weak and non-relativistic ($ v \lesssim 10^{-4}c$, and $U/c^2 \sim 10^{-6}$), it has received little direct experimental verification in the strong-field regime,  which leaves room for alternative theories of gravity that reduce to general relativity in the weak-field limit. Additional evidence supporting general relativity has come from observations of binary pulsar systems which measure several Keplerian and post-Keplerian parameters including the decay of the orbital period of the binary due to binding energy lost in the form of gravitational waves.  The parameters of some alternative theories which predict dipole radiation in addition to the quadrupole radiation predicted by general relativity have been constrained by observations of the neutron-star-white-dwarf systems J1141–6545 and J1738+0333 to an even higher degree than the bounds from the Cassini test in the solar system \cite{EspositoFarese:2004tu, Freire11072012}.

  Research in alternative theories of gravity has been traditionally fueled in large part by the long standing theoretical effort to formulate a quantum theory of gravity, which promises to shed light on big mysteries such as the origin of the universe and the singularities hidden deep within the event horizons of black holes.  A unified theory of gravity and quantum mechanics is necessary because the general relativistic picture of spacetime as a Riemannian manifold breaks down at the length scale $\ell_P \sim 10^{-33}$ cm where quantum fluctuations of the gravitational field become important. However, deviations from classical relativity may be exhibited in the strong-field regime even at macroscopic length scales.  For instance, deviations from Einsteinian gravity at large scales may account for the dark energy content in the universe and active research includes using observational data to constrain models which parameterize such deviations in structure formation models \cite{Tsujikawa:2010zza}.  Furthermore, models with additional scalar fields play a particularly important role in modern theories of cosmology by providing the muscle for the inflationary epoch.

    Among the many alternative theories of gravity, a particularly well studied and well established theory is scalar-tensor (ST) gravity, wherein the agents of the gravitational force are a scalar and a tensor field. ST theories originate from the work of Jordan \cite{Jordan}, Fierz \cite{Fierz},  Brans and Dicke \cite{BransDicke}, who sought a way to incorporate Mach's principle into a covariant theory of gravity.  They found that a dynamical gravitational constant built from a scalar field $\phi \sim G^{-1}$ acted as a way to influence local physics through large scale phenomena, while upholding the weak equivalence principle and maintaining general covariance. The field equations of the theory they developed proved to be equivalent to a sector of the dimensionally reduced five-dimensional field theory designed by Kaluza \cite{Kaluza} as an attempt to unify gravity and electromagnetism.  The emergence of a scalar field(s) is now known to be a generic property of dimensionally reduced higher-dimensional models such as string theory, making the study of gravitational theories with additional scalar fields a very active research field.  
    
    % reason to look at non-stationary scalar field systems
Hawking \cite{hawking1972} has shown that stationary solutions of Brans-Dicke type scalar tensor field theories must also satisfy the Einstein field equations and that the general stationary solutions are the Kerr family of solutions coupled to a constant background scalar field. Building on this,   Sotiriou and Faraoni \cite{Sotiriou:2011dz}  extended Hawking's  proof to include more general ST theories including $f(R)$ theories and relaxed the need for a symmetric collapse. However, the proof fails to hold for certain non-stationary cosmological scalar fields, such as those which asymptote to solutions with linear time dependence at large radii \cite{Jacobson:1999vr}, leading to observable deviations from general relativity in the orbital decay of super-massive black hole binaries \cite{Horbatsch:2011ye}.  Although constraints on such theories do exist, for generality we choose to leave the spacetime dependence of the background scalar field unspecified and we make no restrictions on the behaviour of the scalar potential acting as the cosmological function other than convexity. We then consider a stationary background scalar field configuration as a special case.

 \subsection{Compact binary problem}       
With the promise of gravitational wave observatories directly probing general relativity in the strong-field regime  through gravitational wave (GW) signals, researchers have begun to investigate the dynamics of compact binary systems like binary black holes and neutron stars within various alternative theories of gravity. 

%numerical shit
Scalar-tensor theories benefit from a well-posed Cauchy formulation, enabling numerical relativity simulations to explore strong-field gravitational effects  \cite{Healy:2011ef, Barausse:2012da}.   Although post-Newtonian and linearized theory results indicate that \emph{binary} black holes in ST theory are indistinguishable from their GR counterparts \cite{Will:1989sk,Yunes:2011aa,Mirshekari:2013vb} just like the isolated black holes in the general proofs of Hawking and Faraoni and Sotiriou \cite{hawking1972,Sotiriou:2011dz}, if the scalar field is made time dependent or given some inhomogeneity  through an external mechanism like a potential, the binary also emits dipolar radiation while accreting the scalar field. Healy and collaborators \cite{Healy:2011ef} used a Mexican hat type potential to induce significant non-uniformity in the scalar field to investigate the gravitational and scalar radiation profiles from merging binary black holes in ST theory. Binary neutron star systems, on the other hand, are more sensitive to the presence of the scalar field and display new phenomena that are qualitatively different from predictions of GR.  Isolated neutron stars undergo a phenomenon known as spontaneous scalarization \cite{Damour:sponscalar}, where they develop a sudden scalar charge when they previously had none.   Spontaneous scalarization occurs when a non-zero value of the scalar field inside the star becomes energetically favorable over the zero field configuration for certain values of the ST parameter $\omega$. In the non-linear binary merger regime, \cite{Barausse:2012da} have shown that spontaneous scalarization occurs in much the same way when the merging neutron stars are sufficiently close. The same group has shown that the gravitational wave signals from binary neutron stars may provide constraints on the parameters of ST gravity if one of the companion stars becomes scalarized prior to the time when the gravitational waves enter the detectors sensitivity band \cite{Sampson:2014qqa}. 

% post newtonian 
When the binary objects are of similar size and in the early stages of the inspiral, the fields are relatively weak and the compact bodies move with small orbital velocity relative to the speed of light. Typical examples of these systems include binary neutron stars and stellar mass black holes which are several hundred Schwarschild radii apart.  In this regime, post-Newtonian theory provides the framework for computing the orbital motion and gravitational wave signals. %expanding the fields perturbatively in powers of spatial velocity in the region close to the binary and in power of Newton's in the radiation zone.  
Starting with the work of Wagoner \cite{Wagoner:1970}, researchers have used the post-Newtonian approximation to derive equations of motion for slowly inspiralling binary systems emitting gravitational waves using the field equations of ST theory   \cite{Will:1989sk,Damour:1996ke, Damour:1998jk, Brunetti:1998cc,Mirshekari:2013vb}. These works express the period decay formula in terms of Brans-Dicke parameters and use it to place constraints on the theory through observations in the strong field regime.  The post-Newtonian waveforms in ST theory have been shown to reduce to the general relativity waveforms for binary black holes to second post-Newtonian order in the radiation sector \cite{Lang:2013fna,Lang:2014osa}. Recent interest in cosmological theories with light scalar fields \cite{Wetterich:1994bg,Frieman:1995pm} and the existence of floating orbits \cite{Cardoso:2011xi}, motivated Alsing and collaborators \cite{Alsing:2011er} to explore the post-Newtonian regime of compact binaries with a massive scalar field leading to an explicit expression for the decay of the binary's orbital period in massive Brans-Dicke theory.

 For compact binaries with largely disparate mass parameters,  such as a stellar mass $(m \sim M_{\odot}$) object orbiting a super-massive black hole in the centre of a galaxy $M_{\text{SMBH}} \sim 10^{6} M_{\odot}$, the binary's orbital velocity may enter the highly relativistic regime where post-Newtonian methods no longer work. Instead, one must solve the perturbed Einstein field equations without any recourse to slow-motion expansions. Yunes and collaborators  \cite{Yunes:2011aa} used a test particle approximation to calculate the gravitational wave energy flux coming from an EMRI system consisting of a small body orbiting a Kerr black hole to examine whether low-frequency gravitational wave detections may provide constraints on the Brans-Dicke parameters.  They showed that for massless scalar fields, the constraints are notably worse than those provided by solar-system tests (dephasing of GWs is weakly effected by changes in the parameters), but that massive scalar fields can provide significant constraints \cite{Cardoso:2011xi}.
 
The main motivation for
this work is to provide an additional framework to constrain these theories in the EMRI regime via the effects of the self-force.   The prospect is quite good for nontrivial configurations of the scalar field, due to the presence of new local terms and of several new non-diagonal tail terms that appear in the equations of motion. We will also show that constraints may  be promising even when the scalar field configuration is the trivial one, as in the black hole scenario, due to the existence of a scalar component to
the self-force in addition to the gravitational component. The framework will also serve as a means to dynamically study the floating orbits which arise in motion of particles around spinning black holes in scalar-tensor theories. 
 
	Floating orbits are a byproduct of the superradiant instability displayed by rotating black holes with massive external scalar fields where the net energy flux at infinity is entirely due to the rotational energy extracted from the spinning black hole, halting the inspiral of the small body; i.e., $  \dot{E}_{\mathscr{J}^+} = -  \dot{E}_{r_+}$, implying  $\dot{E}_{\text{orb}}= 0$. In this scenario,  scalar perturbations in the superradiant regime with frequencies close to the mass of the scalar field create resonances which lead to large amounts of negative scalar flux down the horizon balancing the flux of gravitational radiation at null infinity.  The mass of the scalar field is necessary to act as a reflecting wall which traps the radiation in order to create quasi-bound states of the field. The arguments for the existence of floating orbits rely heavily on an adiabatic approximation in which the orbit of the small body is modeled as a sequence of quasi-circular geodesic orbits which get progressively smaller in radius as orbital binding energy is released in the form of gravitational waves.  Alternatively, the power radiated by the particle causing its inspiral may be written in terms of the work done by the dissipative piece of the radiation reaction self-force per unit time. The self-force also contains a conservative component acting on orbital timescales which may effect the stability of the floating orbits. A full calculation of the self-force in the EMRI problem within ST theories is necessary to study how floating orbits arise dynamically and to assess their stability, which is a significant motivator for the present work in addition to the general study of motion in alternative theories.  
 
    The leading order equations of motion for small compact objects in ST theories were given by Gralla \cite{Gralla:2010cd}, who used the Bianchi  identity and various universal scaling relations for the fields in the region outside a small body of mass $M$ and charge $Q$ to map the near-zone Coulomb-type behaviour of the perturbed fields to a set of effective distributional sources. With this he was able to rigorously show that the linearized field equations for a small extended body in Einstein frame ST theory reduced to the point particle form
  \begin{subequations}\begin{align}
 \delta  G_{\a\b} -  8 \pi \delta T_{\a\b} &= 8 \pi \int M(\tau) u_\a u_\b \, \delta_4(x,z(\tau)) d \tau, \\ \delta \Box \Phi &= - 8 \pi \int Q (\tau)\, \delta_4(x,z(\tau))  d\tau,
 \end{align} \end{subequations}
 where $\delta$ is the perturbation operator, $G_{\a\b}$ is the Einstein tensor, $T_{\a\b}$ is the stress-energy tensor of the bulk scalar field, and $\Box:=g^{\a\b}\nabla_\a \nabla_\b$ is the covariant wave operator in the background spacetime. 
 Gralla also noted that the charge $Q$ of the small body is not constrained by any evolution equation inherent to ST theory; it must be prescribed from the model describing the internal structure of the object.  He went on to show that the mass and charge of the small body obey surface integral relations similar to the ADM expressions \cite{Gralla:2013rwa}.  The foundational treatment given by Gralla set the stage for the simultaneous works of Zimmerman and Poisson \cite{Zimmerman:2014uja} and Linz and collaborators \cite{Linz:2014vja}, who overcame the technical challenge of solving the coupled field equations and  derived equations of motion for the coupled self-force and regularization parameters in the scalarvac and electrovac spacetimes of GR. These derivations rely on the Detweiler-Whiting axioms to identify and remove the divergence introduced by the point-particle. In vacuum GR, self-consistent self-force derivations using finite sized bodies and matched-asymptotic expansions performed by Gralla and Wald \cite{gralla-wald:08}, and also Pound \cite{pound:10a}, have reproduced the results of axiomatic point-particle derivations. We expect that our axiomatic derivation will also agree with future non-vacuum self-consistent computations. % \cite{Gralla:2010cd} when extended beyond leading order. %This regularization method is axiomatic in the sense that it the removes the divergence introduced by the point-particle by assuming the Detweiler-Whiting procedure is valid. In vacuum general relativity, the axiomatic procedure adopted here is supported through the calculations of Gralla and Wald \cite{gralla-wald:08}  and also Pound \cite{pound:10a}, who consider finite bodies and use matched-asymptotic expansions to derive the self-force and we expect the same to be true if Gralla's computations \cite{Gralla:2010cd} were to be extended beyond leading order. 
 
   Here we use the technology developed in the aforementioned papers to compute the self-force in generic massive scalar-tensor theories and we find that the resulting  equations of motion are substantially different from those in vacuum GR due to the presence of the additional scalar field.  

The main results of this work are the equations of motion for a small body due to the coupled self-force in a general ST theory in the Einstein frame.  We first give the equation of motion for a background scalar field with arbitrary time dependence and a general potential function.  For this we find that the particle's charge can be expressed in terms of the ST parameters in such a way as to resemble the scalarvac self-force plus additional corrections. We then analyze the asymptotically flat black hole scenario where the scalar field is constant and the potential and its first derivative are zero.  In all cases we find that the equation of motion can be decomposed into the form \[ M a^\m = F^\m_{0} + F^\m_{\text{L}} + F^\m_{\text{tail}} ,\]
where $  F^\m_{0} $ is the force resulting from the gradient of the background potential, $F^\m_{\text{L}}$ is the local contribution to the self-force which is built from background quantities evaluated on the world line, and  $F^\m_{\text{tail}}$ is the non-local contribution to the self-force which takes the form of a time integral over the particle's past history.  The absence of the coupling in the asymptotically flat, constant scalar field black hole case leads to a self-force which is simply the sum of the gravitational and scalar self-forces. 

 The paper begins in Sec~\ref{sec:scalar-vac} with a derivation of the perturbed field equations in the Einstein frame for the general ST theory of a point particle.  In Sec~\ref{sec:notation-review}, we review the condensed index notation presented in \cite{Zimmerman:2014uja} and introduce several quantities to facilitate the local expansion of the perturbed fields. Next, in Sec~\ref{sec:reg-fields} we derive an expression for the regular field and compute its gradient as required  for the self-force. Finally, we derive the self-force equation of motion for a point-particle in ST theory (Sec.~\ref{sec:eom}) and specialize it to a black hole in an asymptotically flat spacetime (Sec.~\ref{sec:bh}).

%%%%%%%%%%%%%%%%%%%%%%%%%%%%%%%%%%%%%%%%%%%%%%%%%%%
% SECTION : ST THEORY ACTION AND EQS
\section{The scalar-tensor theory of a single scalar field} \label{sec:scalar-vac}
%%%%%%%%%%%%%%%%%%%%%%%%%%%%%%%%%%%%%%%%%%%%%%%%%%%
The quantity of information on ST theories is substantial, but here we limit ourselves to introducing elementary quantities and concepts of ST theory needed for the purposes of the self-force derivation. A detailed review of the theory may be found in \cite{STbook}.  

The development here is to start with the action in the Jordan frame and then perform a conformal transformation to the Einstein frame where the field operator is the Einstein tensor and the scalar field contributes only as matter content through the bulk stress-energy tensor. We then vary the action with respect to the Einstein frame fields to obtain the field equations.  This allows us to express the perturbed field equations in manifestly hyperbolic form after a suitable gauge transformation. From there we can express the perturbed field equations in a form where we can  directly apply the multi-field methods of \cite{Zimmerman:2014uja}.
 
The action for a generic scalar-tensor theory in the Jordan frame has the form 
\begin{equation}
S_J = \frac{1}{16\pi} \int  \Bigl( a(\bar{\phi}) \bar{R} - b(\bar{\phi}) \bar{g}^{\mn} \bar\nabla_\m \bar{\phi} \bar\nabla_\n \bar\phi - 2 c(\bar\phi) \Bigr)  \sqrt{- \bar{g}}  \, d^4 x + S_{\text{M}}(\Psi_{\text{M}},\bar{g}^{\mn}),
\end{equation} 
where $\Psi_{\text{M}}$ collects the fields responsible for the matter content of the theory, $a$, $b$ and $c$ are field-dependent ST parameters, and the overbar indicates that the quantity is taken with respect to the Jordan frame.  Without loss of generality, we can redefine the scalar field $ a(\bar\phi) \rightarrow \bar\phi$. After the field redefinition we see that  the most general ST theory contains two free functions: a coupling function (typically written as $\omega(\phi)$) which precedes the kinetic term and a cosmological function which enters without any explicit powers of the scalar field.  The coupling function is responsible for the spontaneous scalarization phenomenon mentioned in the introduction where the binary develops a sudden scalar charge \cite{Damour:1998jk}. The effects of the cosmological function include providing the scalar field with mass and playing the role of the cosmological constant. The potential also contributes corrections to the mass evolution of scalar particles in curved spacetime as we shall see. 

  In the EMRI scenario of vacuum general relativity, the matter content is a small body which is approximately described by a structureless point-particle of mass $m$. In the ST theory of material objects, however,  the inertial mass and structural properties of the small body will generally be influenced by the scalar field due to the variability of Newton's gravitational constant. This creates a need to introduce a new parameter which parameterizes the sensitivity of the body's binding energy on the background scalar field.  The sensitivity of neutron stars will depend on their radius and equation of state but black holes all have the same sensitivity. % s=1/2 in BD theory.  
The dependence on the scalar field is incorporated into the point-particle model by allowing the mass to vary with the scalar field \cite{Eardley:1975}. The point-particle action for the theory is thus given by 
\begin{equation}
S_{\text{M}} = - \int  m(\bar\phi)\, d \bar\tau
\end{equation}
in the Jordan frame.
Although the Jordan frame benefits from explicit conservation of stress-energy $\nabla_\a T^{\a\b}_J =0$, the field equations in the Jordan frame do not lend themselves to a direct application of the methods of Zimmerman and Poisson \cite{Zimmerman:2014uja}.  For this, we must   transform the fields into the Einstein frame.  Performing the conformal transformation \cite{Damour:1996ke}
\begin{subequations}
\begin{align}
g_{\mn} &= a(\bar\phi) \bar g_{\mn},\\
\chi(\bar \phi) &= \int \left( \frac34 \left( \frac{a'(\phi)}{a(\phi)}\right)^2 + \frac12 \frac{b(\phi)}{a(\phi)}\right) d \phi, \\
A(\chi) &= a^{-1/2}(\bar\phi), \\
F(\chi) &=  \frac{ c(\bar\phi) }{a^{2}(\bar\phi) },
\end{align}
\end{subequations}
we are led to the Einstein frame action 
\begin{equation}
S_E = \frac{1}{16 \pi} \int \left( R - g^{\mn} \nabla_\m \chi \nabla_\n \chi - 2 F(\chi) \right) dV - \int A(\chi) m(\chi) d \tau,
\end{equation}
where we have introduced the covariant volume element in the Einstein frame $dV:=\sqrt{-g}d^4 x$.
A variation with respect to the metric yields the field equations in the Einstein frame 
\begin{equation}
G_{\a\b} = 8 \pi \left( T_{\ab}^{\text{bulk}} + T_{\ab}^{\text{pp}} \right),
\end{equation}
where 
\begin{equation}
T_{\mn}^{\text{bulk}} = \frac{1}{8\pi} \left( \nabla_\m \chi \nabla_\n \chi - \frac12g_{\mn} \left( \nabla_{\lambda} \chi \nabla^\lambda \chi +2 F \right)\right)
\end{equation}
is the stress-energy of the bulk scalar field
and 
\begin{equation}
T^{\text{pp}}_{\mn} = \int A(\chi) m(\chi) u_\m u_\n \delta_4(x,z) d\tau
\end{equation}
is the stress-energy of the point particle. 
Using the Einstein field equation, we find that the Ricci tensor in terms the matter field reads
 \beq R_{\a\b} = \nabla_\a \chi \nabla_\b \chi + F g_{\a\b}. \eeq
Varying the action with respect to the scalar field produces the scalar wave equation
\begin{equation}
\Box \chi - F'(\chi) = 8 \pi \int \frac{ d ( A m ) }{d \chi } \delta_4 (x,z) d \tau
\end{equation}
governing the evolution of $\chi$, where $\Box := g^{\mn} \nabla_\m \nabla_\n$ 
is the covariant wave operator and $F'(\chi) := dF / d\chi$.  The presence of the point-particle creates perturbations of the fields around their background values. If we denote the background fields (full fields taken at $m=0$) by $\chi(0):=\Phi$ and $g_{\mn}(0) := \tilde{g}_{\mn}$, the perturbed fields are given by
\begin{subequations}
\begin{align}
f &:= \chi - \Phi, \\
h_{\mn} &:= g_{\mn} - \tilde{g}_{\mn}.
 \end{align}
 \end{subequations}
In what follows we always work with either the metric perturbation or the background metric, so we drop the tilde notation and refer to the background metric simply as $g_{\a\b}$. It is convenient to work with the trace-reversed metric perturbation  \[ \gamma_{\a\b} := h_{\a\b} - \frac12 g_{\a\b} h, \] where $h= g^{\a\b} h_{\a\b}$. The benefit of the trace-reversed metric is that it has zero divergence in the Lorenz gauge.  Here we require the metric perturbation to satisfy the one-parameter family of gauge conditions \[ \nabla_{\a} \gamma^{\a\b} = 2 \lambda f \nabla^{\b} \Phi \] in order to put the field equations in weakly hyperbolic form.  The advantage of this gauge over the standard Lorenz gauge is that it eliminates the derivative coupling when $\lambda=1$.
 The perturbed scalar field evolves according to the linearized  equation
 \begin{equation}
 \Box f + N^{\cdot}_{\ | \a\b} \gamma^{\a\b} + N^{\cdot}_{\ \vert \cdot} f = - 4 \pi \rho  
 \end{equation}
 where 
 \begin{subequations}
\begin{align}
 N^{\cdot}_{\ | \a\b} &= - \left(\nabla_\a \nabla_\beta \Phi - \frac12 F' g_{\a\b}\right), \\
 N^{\cdot}_{\ \vert \cdot} &= -\left( 2 \lambda \nabla_\gamma  \Phi \nabla^\gamma \Phi + F''\right),
 \end{align}
 \end{subequations}
play the role of external potentials and the scalar source is given by
\begin{align}
\rho &= - 2 \int m(\Phi) A(\Phi) \left( \frac{A'(\Phi)}{A(\Phi)} +\frac{m'(\Phi)}{m(\Phi)} \right) \delta_4(x,z) d\tau \\  
&:= - 2 \int   m(\Phi) A(\Phi) \alpha(\Phi) \delta_4(x,z) d\tau  .
\end{align}
The use of the ``dot'' and ``bar'' notation is borrowed from \cite{Zimmerman:2014uja} and will be reviewed in a later section. The perturbed Einstein equation has the form \[ \delta G_{\a\b} = 8 \pi \left ( \delta T_{\a\b}^{\text{bulk}} +  t_{\a\b} \right), \]
where the perturbed Einstein tensor is given by
\begin{align}
2 \delta G_{\a\b} =  - \Box \gamma_{\a\b} &+ 2 \nabla_{ (\a\vert } \nabla_{\lambda} \gamma^{\lambda}_{ \ \vert \b )} 
- g_{\a\b} \nabla_\c \nabla_\d \gamma^{\cd} - 2 R^{\c \ \d }_{\ \a \ \b} \gamma_{\cd}  \\ \no &+  2 R_{\m ( \a} \gamma^{\m}_{\ \b)} + g_{\a\b} R^{\cd}\gamma_{\cd} - R \gamma_{\ab} .
\end{align}
The perturbed Einstein tensor is sourced by  the perturbed stress-energy of the scalar field
\begin{align}
8 \pi \delta T^{\text{bulk}}_{\a\b} = & 2 \nabla_{ ( \a } f \nabla_{\b )} \Phi - g_{\a\b}\nabla_{\lambda}\Phi \nabla^{\lambda} f - g_{\a\b} F' f + \frac12 g_{\a\b} \gamma^{\cd}\nabla_\c \Phi \nabla_\d \Phi  \\ \no &- \left( \frac12 \nabla_\m \Phi \nabla^\m \Phi + F \right)\gamma_{\ab} + \frac12 g_{\ab} F g^{\m\n} \gamma_{\m\n}
\end{align} 
in the bulk and the stress-energy of the point-particle 
\beq 
t_{\a\b} = \int A(\Phi) m(\Phi)  u_\a u _\b \delta_4(x,z) d \tau
\eeq
on the world line.
  Putting everything together, 
the perturbed Einstein field equation takes the form
\beq
\Box \gamma^{\a\b} +  M^{\a\b}_{\ \ \ \vert\cdot \gamma} \nabla^{\gamma}f  + N^{\a\b}_{\ \ \ \vert \c\d} \gamma^{\cd} + N^{\a\b}_{\ \ \ \vert \cdot} f= -16 \pi t^{\a\b},
\eeq
where 
\begin{subequations} 
\label{MN-scalar1} 
\begin{align} 
M^{\alpha\beta}_{\ \ \ |\cdot\mu}&:= 2(1-\lambda) \bigl( 
\delta^{\alpha}_{\ \mu} \nabla^\beta \Phi 
+ \delta^{\beta}_{\ \mu} \nabla^\alpha \Phi 
- g^{\alpha\beta} \nabla_\mu \Phi \bigr), \\ 
N^{\alpha\beta}_{\ \ \ |\gamma\delta}&= 
2 R^{\alpha\ \, \beta}_{\ (\gamma\ \delta)} 
- \delta^\alpha_{\ (\gamma} \nabla^\beta \Phi \nabla_{\delta)} \Phi 
- \delta^\beta_{\ (\gamma} \nabla^\alpha \Phi \nabla_{\delta)} \Phi, \\ 
N^{\alpha\beta}_{\ \ \ |\cdot} &= -2 \bigl[ 
(1-\lambda) g^{\alpha\beta} F'  
+ 2\lambda \nabla^{\alpha} \nabla^\beta \Phi \bigr].  
\end{align} 
\end{subequations} 
Note that $M^{\alpha\beta}_{\ \ \ |\cdot\mu}=0$  for the choice of gauge parameter given by $\lambda=1$, which removes all derivative coupling from the field equations.  

\section{Regular field}
\subsection{Field equations in compact form, Hadamard expansion, and regular/singular field decomposition}\label{sec:notation-review}
The two coupled field equations can be  jointly solved  using the condensed index notation introduced by Poisson and Zimmerman \cite{Zimmerman:2014uja}, which we briefly review here.  In the condensed index notation, we use an upper-case latin index to denote the tensorial type of the field. In the present context, upper case indices will denote either a 2-tensor $A=\a\b$ as a pair of indices, or a 0-tensor  (scalar) $A=\cdot$ as the absence of tensor indices. The two fields $f$ and $\gamma_{\a\b}$ are combined into the field doublet $\psi_A = \{f,\gamma_{\a\b} \} $. Similarly, the world line  sources $t_{\a\b}$ and $\rho$ are collected into the doublet \[ \mu^A = \int   g^A_{\ M}(x,z) q^M(\tau) \,  \delta_4(x,z) d\tau \] where 
\begin{equation} 
q^A := \left\{ 
\begin{array}{l} 
4 m(\Phi) A(\Phi) u^\alpha u^\beta, \quad \quad  A= \a\b \\  
-2 m(\Phi) A(\Phi) \alpha(\Phi), \quad\, A= \cdot
\end{array}   
\right. ,
\end{equation}
and $g^A_{\ M}(x,z) $ defines the parallel propagator which transports tensors at $x$ to $x'$ \cite{PPV} 
\begin{equation} 
g^A_{\ B'}(x,x') := \left\{ 
\begin{array}{l} 
g^{(\alpha}_{\ \gamma'}(x,x') g^{\beta)}_{\ \delta'}(x,x'), \quad \quad A= \a\b,\, B'= \c'\d' \\ 
1,   \quad  \qquad\qquad\qquad\quad\quad\,\,\, A=B=\cdot
\end{array} 
\right. .
\label{gAB_def} 
\end{equation} 
Note that the parallel propagator is zero off the diagonal. 
In the condensed notation, both perturbation equations are combined into the single equation 
\beq\label{eq:PsiAB-eq} \Box \psi^A + M^A_{\ B \lambda} \nabla^\lambda \psi^B + N^{A}_{ \ B} \psi^B = - 4 \pi \mu^A. \eeq
As \eqref{eq:PsiAB-eq} is a linear wave equation, its solution can be represented as a Green function contracted with the source
\beq\label{eq:PsiAB-sol}
\psi_{A} = \int G_{AB'} \mu^{B'} d V',
\eeq
where $G_{AB'}$ obeys the wave equation \begin{equation} 
\Box G^A_{\ B'}(x,x')  + M^A_{\ B\mu} \nabla^\mu G^B_{\ B'} (x,x') 
+ N^A_{\ B} G^B_{\ B'}(x,x')  = -4\pi g^A_{\ B'} \delta_4(x,x') 
\label{metaGreen} 
\end{equation} 
and is chosen to satisfy retarded boundary conditions, as all radiation is purely outgoing. The principal parts of the off-diagonal Green functions are solutions to homogeneous equations; e.g. $\Box G^{\cdot}_{\ \ \vert \a\b} =0 $, whereas the principal parts of the diagonal Green functions have distributional sources.   For all $x$ in the convex normal neighbourhood of a base point $x'$, the retarded Green function can be cast in the Hadamard form 
\begin{equation} 
G^A_{\ B'}(x,x') = U^A_{\ B'}(x,x') \delta_+(\sigma) 
+ V^A_{\ B'}(x,x') \Theta_+(-\sigma),
\label{Hadamard} 
\end{equation} 
in which $\sigma(x,x')$ is Synge's biscalar, $\delta_+(\sigma)$ 
and $\Theta_+(-\sigma)$ are the Dirac and Heaviside
distributions supported in the future of $x'$, and $U^A_{\ B'}(x,x')$
and $V^A_{\ B'}(x,x')$ are smooth bitensors.  The coefficients $U^A_{\ B'}(x,x')$
and $V^A_{\ B'}(x,x')$ are recursively determined using the field equations \cite{PPV}. 
	
In the normal neighbourhood of the world line $z(\tau)$ the retarded solution takes the form of a local leading-order $r^{-1}$ piece, and a pair of tail integrals given by
\begin{equation} 
\psi^A(x) = \frac{1}{r} U^A_{\ B'}(x,x') q^{B'}(u) 
+ \int_{\tau_<}^u V^A_{\ M}(x,z) q^M(\tau)\, d\tau
+ \int_{-\infty}^{\tau_<} G^A_{\ M}(x,z) q^M(\tau)\, d\tau ,
\label{psi_ret} 
\end{equation}
where $u$ is the retarded time at the point $z(u)$ where a past directed null ray %having affine parameter $r$ 
starting from $x$ intersects the world line, $v$ is the advanced time of a point connecting $x$ to the world line by a future directed null ray, $r$ is the retarded distance from $x$ to $z(u)$,  and $\tau_{<}$ is the proper time where the world line intersects the convex normal neighbourhood of $x$ \cite{PPV}.  

The near-zone behaviour of the solution \eqref{psi_ret} exhibits a Coulomb-type $1/r$ behaviour which leads to a singularity at the location of the particle $r=0$. 
The divergence is regularized using the Detweiler-Whiting prescription. The method involves constructing a singular field $\psi_{\sf S}$ and subtracting it from the retarded field, i.e., $\psi_{\sf R} := \psi - \psi_{\sf S}$. The singular field is built from a ``singular'' Green function which vanishes in the causal future and past of the world point, is symmetric in its arguments, and solves the inhomogeneous wave equation \cite{PPV}.  The Detweiler-Whiting regular field is found to be 
{\small \begin{align} \label{psi_reg} 
\psi^A_{\sf R}(x) &= \frac{1}{2r} U^A_{\ B'}(x,x') q^{B'}(u) 
- \frac{1}{2r_{\rm adv}} U^A_{\ B''}(x,x'') q^{B''}(v) 
+ \int_{\tau_<}^u V^A_{\ M}(x,z) q^M(\tau)\, d\tau
 \\ \nonumber & \quad \mbox{} 
+ \frac{1}{2} \int_u^v V^A_{\ M}(x,z) q^M(\tau)\, d\tau 
+ \int_{-\infty}^{\tau_<} G^A_{\ M}(x,z) q^M(\tau)\, d\tau, 
\end{align} }
where $r_{\rm adv}$ is the advanced distance from $x$ to the world line and $v$ is the advanced time at that point. 

%%%%%%%%%%%%%%%%%%%%%%%%%%%%%%%%%%%%%%%%%%%%%%%%%%%%%%%
\subsection{Regular field in Fermi normal coordinates}\label{sec:reg-fields}
%%%%%%%%%%%%%%%%%%%%%%%%%%%%%%%%%%%%%%%%%%%%%%%%%%%%%%%
In this section we derive expressions for the regular fields in local coordinates. We choose to adopt Fermi normal coordinates (FNC) adapted to the world line of the body.  To construct a system of FNC of a point $x$ centred on the world line, we choose a point on the world line $\bar{x}:=z(t)$, where $t$ is the proper time at the point. We then locate the unique spacelike geodesic which is orthogonal to the world line at $\bar x$  that connects the two points $x$ and $\bar x$  in the normal neighbourhood.   Along the connecting geodesic one can define a tangent bisector $\sigma^{\bar\a}(\bar x,x) $ which is orthogonal to the four velocity $u^{\bar\a}$ of the world line. Here the barred index indicates that the quantity transforms tensorially at the barred point. The defining relations for the coordinate system are  therefore $\bar{x}^0 = t$, $\bar{x}^a = -e^a_{\ \bar\a} \sigma^{\bar\a}$ and $\sigma^{\bar\a} u_{\bar\a} =0$, where $e^a_{\bar\a}$ is a spatial triad that is Fermi-walker transported along the world line. The geodesic distance from $\bar x$ to $x$ running along the spatial curve is given by $s^2 = x_a x^a = 2 \sigma$, which is perturbatively small relative to the curvature scale. 

 To compute the regular field in FNC, the retarded/advanced time dependencies in the expression \eqref{psi_reg} coming from the terms $U^A_{\ B'}(x,x') q^{B'}(u)$ 
and $ U^A_{\ B''}(x,x'') q^{B''}(v)$  must be translated into dependencies on the Fermi point $\bar{x}$.  This is accomplished  by starting with the Taylor expansion of the direct part of the Hadamard form 
\begin{equation} 
U^A_{\ B'}(x,x') = g^A_{\ A'} \biggl[ \delta^{A'}_{\ B'} 
+ U^{A'}_{\ B'\mu'} \sigma^{\mu'} 
+ \frac{1}{2} U^{A'}_{\ B'\mu'\nu'} \sigma^{\mu'} \sigma^{\nu'} 
+ O(\epsilon^3) \biggr], 
\label{U_taylor1}
\end{equation}  
where 
\begin{subequations} 
\label{U_taylor2} 
\begin{align} 
U^{A'}_{\ B'\mu'} &= \frac{1}{2} M^{A'}_{\ B'\mu'}, \\ 
U^{A'}_{\ B'\mu'\nu'} &= 
-\frac{1}{2} \nabla_{(\mu'} M^{A'}_{\ B'\nu')} 
+ \frac{1}{4} M^{A'}_{\ C'(\mu'} M^{C'}_{\ B'\nu')} 
+ \frac{1}{6} \delta^{A'}_{\ B'} R_{\mu'\nu'},
\end{align} 
\end{subequations} 
about the base point ${x'}$ at the retarded time $u$, and then re-expanding it %the dependence on the retarded/advanced base points 
around the Fermi point $\bar x$ at the present time $t$.   The retarded and advanced distances must also be expanded in powers of the spatial distance $s$ with coefficients being evaluated on the world line at time $t$.  The details of the procedure are spelled out explicitly in Refs.~\cite{PPV, Zimmerman:2014uja}.  

 The regular field in FNC to $\orderof{s}$ written in terms of condensed index notation was derived in \cite{Zimmerman:2014uja} and found to be
 {\footnotesize
\begin{subequations}
\label{psi_fermi} 
\begin{align} 
 \psi^A_{\sf R}(t,x^a) &= -(1-a_c x^c) \dot{U}^A(t) 
+ \frac{1}{3} U^A(t)\, \dot{a}_c x^c 
+ \psi^A[\text{tail}] + O(s^2) 
\label{psi_fermi_a} \\ 
&= -g^A_{\ \bar{A}} \Bigl( \dot{\hq}^{\bar{A}} 
+ \hq^{\bar{B}} U^{\bar{A}}_{\ \bar{B} t} \Bigr) (1-a_c x^c) 
+ g^A_{\ \bar{A}} \biggl( \frac{1}{3} \hq^{\bar{A}} \dot{a}_c 
+ \hq^{\bar{B}} \dot{U}^{\bar{A}}_{\ \bar{B} c} 
+ \dot{\hq}^{\bar{B}} U^{\bar{A}}_{\ \bar{B} c}  
+ \hq^{\bar{B}} U^{\bar{A}}_{\ \bar{B} tc} \biggr) x^c 
\nonumber \\ & \quad \mbox{} 
+ \frac{1}{2} \hq^{\bar{B}} R^A_{\ \bar{B} tc}\, x^c 
+ \psi^A[\text{tail}] + O(s^2), 
\label{psi_fermi_b} 
\end{align} 
\end{subequations}}
where $\hq^{\ab}=q^{\a\b}$, $\hq = \frac12 q$ \cite{Zimmerman:2014uja}, and $\psi^A[\text{tail}]$ denotes the contribution from the chronological past 
\[ \psi^A[\text{tail}](x) = \lim_{\epsilon\rightarrow 0} \int_{-\infty}^{t- \epsilon}  G_{AB} (x,z(\tau))q^B (\tau) d \tau  := \int_{-\infty}^{t_-}  G_{AB} (x,z(\tau))q^B (\tau) d \tau. \] Note that we cut the integral short before taking the limit $x\rightarrow \bar x$ and not the reverse.  We choose to work with $\hq$ because it is the quantity which appears in the background equations of motion $ A m a_\a = {\hq} \nabla_\a \Phi := Q \nabla_\a \Phi$, which we interpret as the charge of the body. Under this definition of the charge, the ``sensitivity'' of the object is $s :=- d\ln M /d \ln \Phi =\frac{\Phi}{A m} Q$. 

Inserting $M^A_{\ B\mu}$ and $N^A_{\ B}$ from
Sec.~\ref{sec:scalar-vac} into
Eqs.~(\ref{U_taylor2}) and (\ref{V_taylor2}) we find that the $U$ part of Hadamard Green function which enters the regular field is given explicitly by  
\begin{subequations} 
\label{grav_scalar_tensorlist-1} 
\begin{align} 
U^A_{\ B\mu} &= 0, \\ 
U^{\alpha\beta}_{\ \ \ |\gamma\delta\mu\nu} &= 
\frac{1}{6} \delta^{(\alpha}_{\ \gamma} \delta^{\beta)}_{\ \delta} 
R_{\mu\nu}, \\ 
U^{\cdot}_{\ |\cdot\mu\nu} &= \frac{1}{6} R_{\mu\nu}. 
\end{align} 
\end{subequations} 
The gradient of the regular field, which yields the force, also includes contributions from the $V$-terms in the Hadamard form.  The $V$-terms are computed with the help of the expansion
\begin{equation} 
V^{A'}_{\ B'} = -\frac{1}{4} \nabla^{\mu'} M^{A'}_{\ B'\mu'} 
- \frac{1}{8} M^{A'}_{\ C'\mu'} M^{C'\ \mu'}_{\ B'} 
+ \frac{1}{2} N^{A'}_{\ B'} 
+ \frac{1}{12} \delta^{A'}_{\ B'} R', 
\label{V_taylor2} 
\end{equation} 
and we find
\begin{subequations} 
\label{grav_scalar_tensorlist-2}
\begin{align} 
V^{\alpha\beta}_{\ \ \ |\gamma\delta} &= 
R^{\alpha\ \, \beta}_{\ (\gamma\ \delta)} 
- \delta^{\alpha}_{\ (\gamma} \nabla^\beta \Phi \nabla_{\delta)} \Phi 
+ \frac{1}{12} \delta^{\alpha}_{\ (\gamma} \delta^{\beta}_{\ \delta)}
    R, \\
V^{\alpha\beta}_{\ \ \ |\cdot} &= -2 \nabla^\alpha \nabla^\beta \Phi, \\ 
V^\cdot_{\ |\alpha\beta} &= -\frac{1}{2} \biggl( 
  \nabla_\alpha \nabla_\beta \Phi - \frac{1}{2} F' g_{\alpha\beta}
  \biggr), \\ 
V^\cdot_{\ |\cdot} &= -\frac{1}{2} \Bigl( 
  2 \nabla_\mu \Phi \nabla^\mu \Phi + F'' \Bigr) 
+ \frac{1}{12} R. 
\end{align} 
\end{subequations} 

The regular scalar field to next-to-leading-order in the Fermi parameter is given by the expression
\begin{equation}\label{eq:regscalar}
f_{\sf R} = - \left(1-a_c x^c\right) \dot{\hq} + \frac16 \hq \left(2 \dot{a}_c + R_{tc} \right) x^c + f[\text{tail}] + \orderof{s^2}.
\end{equation}
To compute the regular metric perturbation we rely on the FNC expansion of the parallel propagator given by
\beq\label{eq:pp}
g^{t}_{\ \bar t} = 1- a_a x^a + \orderof{s^2}, \quad g^t_{\ \bar a} = \orderof{s^2}, \quad  \quad g^a_{\ \bar b} = \delta^a_{\ b} + \orderof{s^2} .
\eeq
 Substituting the expansion for the parallel propagator into Eq.~\eqref{psi_fermi_b}, we find that the regular metric perturbation takes the form
\begin{subequations} 
\label{eq:reggamma} 
\begin{align} 
\gamma^{\a\b}_{\sf R} = &- \dot{q}^{tt} e^{\a}_{\ t} e^{\b}_{\ t} - 2 \dot{q}^{tb} e^{(\a}_{ \ t} e^{\b)}_{ \ b} \\ \no &+ \Bigg[
 e^{\a}_{\ t} e^{\b}_{\ t} \left(3 a_c \dot{q}^{tt} + \frac13 q^{tt} \dot{a}_c + \frac16 q^{tt} R_{tc} \right)+ e^{(\a}_{ \ t} e^{\b)}_{ \ b}\left( 4 \dot{q}^{tb} a_c - q^{tt} R^b_{\ t c t} \right) \Bigg ] x^c \\  \nonumber
 &+ \gamma^{\a\b}[\text{tail}] + \orderof{s^2}.
\end{align}
\end{subequations} 
In Eqs. \eqref{eq:regscalar} and \eqref{eq:reggamma}, the tail terms are defined by the relations
\beq\label{eq:gamma-tail}
\gamma^{\ab} [\text{tail}] := \int_{-\infty}^{t^-} d\tau \, G^{\a\b}_{\ \ \vert \c\d}(x,z) q^{\c\d}(\tau) +    \int_{-\infty}^{t^-} d\tau \, G^{\a\b}_{\ \ \vert \cdot}(x,z) \hat{q}(\tau)
\eeq
and 
\beq\label{eq:f-tail}
f[\text{tail}] := \int_{-\infty}^{t^-} d\tau \, G^{\cdot}_{\ \ \vert \c\d}(x,z) q^{\c\d}(\tau) +    \int_{-\infty}^{t^-} d\tau \, G^{\cdot}_{\ \ \vert \cdot}(x,z) \hat{q}(\tau).
\eeq
The equation of motion is written in terms of the non-trace-reversed metric perturbation $h_{\a\b} = \gamma_{\a\b} - \frac12 g_{\a\b} \gamma$, which we now compute. Using the expansions of the background metric in FNC $g_{tt} = -1-2 a_a x^a + \orderof{s^2}$, $g_{ta} = \orderof{s^2}$, $g_{ab} = \delta_{ab} + \orderof{s^2}$, and the trace of $\gamma_{\a\b}$, $\gamma = \dot{q}^{tt} - \left(a_c \dot{q}^{tt} + \frac13 q^{tt} \dot{a}_c + \frac16 q^{tt} R_{tc} \right)x^c + \orderof{s^2}$, we find 
\begin{subequations} \label{eqs:h-reg}
\begin{align} 
h_{tt}^{\sf R} &= -\frac12 \dot{q}^{tt} + \left( -\frac12 \dot{q}^{tt}a_c + \frac16 q^{tt} \dot{a}_c + \frac12 q^{tt}R_{tc}\right)x^c+ h_{tt}[\text{tail}] + O(s^2), \\ 
h_{ta}^{\sf R} &= \delta_{ab} \dot{q}^{tb} + \frac12 q^{tt} R_{atct} x^c 
+ h_{ta}[\text{tail}] + O(s^2), \\ 
h_{ab}^{\sf R} &= -\frac12 \delta_{ab} \dot{q}^{tt} + \frac12 \delta_{ab} \left(\dot{q}^{tt}a_c +\frac13 q^{tt} \dot{a}_c + \frac16 q^{tt} R_{tc} \right) x^c 
+ h_{ab}[\text{tail}] + O(s^2).
\end{align} 
\end{subequations} 
Using the definition of the background gradient $\nabla_ \a h_{\b\c} = \pd_\a h_{\c\d} - \Gamma^\mu_{\ \b\a}h_{\m\c} - \Gamma^{\m}_{\ \c\a}h_{\b\m}$, and the FNC expansion of the Christoffel symbols $\Gamma^{t}_{ \ t a} = a_a + \orderof{s^2}$, $\Gamma^{a}_{\ tt} = a^a + \orderof{s^2}$, with all other components being higher order, we find

\begin{subequations}
\begin{align} 
\nabla_a h_{tt}^{\sf R} &= \frac12 \dot{q}^{tt} a_a + \frac16 q^{tt} \dot{a}_a + \frac{1}{12} q^{tt}R_{ta} + \nabla_a h_{tt}[\text{tail}]+ O(s), \\ 
 \nabla_t h_{ta}^{\sf R} &= \delta_{ab} \ddot{q}^{ \ tb} + \dot{q}^{tt} a_a + \nabla_t h_{ta}[\text{tail}] + O(s), 
\end{align} 
\end{subequations} 
and 
\begin{equation}
\nabla_a f_{\sf R} = \dot{\hat{q}} a_a + \frac{1}{6} {\hat q}  \left( 2 \dot{a}_a + R_{ta} \right) + \nabla_a f[\text{tail}] + \orderof{s}. 
\end{equation}
 
We get an additional local contribution from the time derivatives of the tail in the normal neighborhood at the current time $t$. Recall that 
\begin{equation}
f[\text{tail}] = \int_{\tau_<}^{t} V^{\cdot}_{\ A} \hat{q}^A d\tau + \int_{-\infty}^{\tau_<} G^{\cdot}_{\ \vert A} \hat{q}^A d\tau
\end{equation} 
which when differentiated gives
\begin{equation}
\nabla_t f[\text{tail}] = V^{\cdot}_{\ \vert \cd} q^{\cd} + V^{\cdot}_{\ \vert \cdot} \hat{q} + \nabla_t f^{(\epsilon)}[\text{tail}] + \orderof{s}
\end{equation}
after using relations \eqref{grav_scalar_tensorlist-2} and \eqref{V_taylor2}. We use the notation $\nabla_t f_{\sf R}^{(\epsilon)}[\text{tail}]$ to denote the part of the tail that's beyond leading order in the FNC expansion, which excludes the local part coming from $V$, which is smooth, and the contribution from $G^\cdot_{\ \vert A}$ at the particle, which is singular.  
%\begin{equation}
% = \lim_{\eps\rightarrow 0}\int^{t-\eps}_{-\infty} G^{\cdot}_{\ \vert A} \, {\hat q}^{A}.
%\end{equation}
 Likewise, we find 
 \begin{align}
 \nabla_t h^{\sf R}_{\ ta} [\text{tail}] &= \Lambda_{ta}^{ \ \ \a\b} \left( {V}_{\a\b\vert \c\d} q^{\cd} + {V}_{\a\b\vert \cdot}{\hat q} \right) + \nabla_t h_{ta}^{(\eps)}[\text{tail}] + \orderof{s} \\ \no
&= \frac12 q^{tt} \dot{\Phi} \nabla_a \Phi  -2 \hat{q}\nabla_t \nabla_a \Phi + \nabla_t h_{ta}^{(\eps)}[\text{tail}] + \orderof{s},
 \end{align}
 where $\Lambda_{\a\b\c\d} := \frac12 (g_{\a\c}g_{\b\d} + g_{\a\d}g_{\b\c} - g_{\a\b}g_{\c\d})$ reverses the trace.
%%%%%%%%%%%%%%%%%%%%%%%%%%%%%%%%%%%%%%%%%%%%%%%%%%%%%%%
\section{Equation of motion}\label{sec:eom}
%%%%%%%%%%%%%%%%%%%%%%%%%%%%%%%%%%%%%%%%%%%%%%%%%%%%%%%
Just as we derived the field equations from the stationarity of the action under first variation, we find the equation of motion in the background by varying the Einstein frame point particle action in the background geometry
\begin{equation}
S_{\text{pp}} = - \int A(\Phi) m(\Phi) \, d \tau, 
\end{equation}
which gives the equation
\begin{equation}
 \left( m  A \right)  a^\a = - \left( A m \a \right) w_{\a}^{ \ \b} \nabla_\b \Phi  := Q w_{\a}^{ \ \b} \nabla_\b \Phi ,
 \end{equation}
 where \[ w_\a^{\ \b} := \left( \delta_\a^{\ \b} + u_\a u^\b\right) \] projects in the directions orthogonal to $u_\a$. The evolution of the particle's mass is dictated by the differential equation 
 \begin{equation}
 \frac{d (m A) }{d\tau} = - Q u^\m \nabla_\m \Phi. 
 \end{equation}
 Likewise, we find the equation of motion in the perturbed geometry by varying the first-order perturbed point particle action 
\begin{equation}
S^{1}_{\text{pp}} =  \frac18 \int q^{\m\n} h_{\m\n} \,d \tau + \int \hat{q} f \, d \tau
\end{equation}
with respect to the coordinate $z^\a$. We find 
\begin{equation}
\delta S^1_{\text{pp}} = \int d \tau\, \delta z^\a \Biggl [ \left(a_\a + w_{\a}^{ \ \lambda}\nabla_\lambda \right) \left( {\hat q} f + \frac18 q^{\mn}h_{\mn} \right) - \frac18 \left( w_{\a}^{ \ \lambda} \frac{D}{d\tau} + 2 g^{\lambda\rho} a_{(\a}u_{\rho)} \right) \frac{ \delta q^{\mn} }{\delta u^\lambda} h_{\mn} \Biggr]
\end{equation} 
Working in FNC where 
\begin{subequations}\begin{align}
q^{\mn} &= q^{tt} u^\m u^\n, \\
\dot{q}^{\mn} &= \dot{q}^{tt} + 2 q^{tt} a^{(\m}u^{\n)}, \\
\frac{\d q^{\mn}}{\d u^\lambda} &= 2  q^{tt} \d^{(\m}_{ \ \lambda}u^{\n)} ,\\
\frac{D}{d \tau} \frac{\d q^{\mn}}{\d u^\lambda} &= 2 \left( \dot{q} ^{tt} \d^{(\m}_{ \ \lambda}u^{\n)} + 2  q^{tt} \d^{(\m}_{ \ \lambda}a^{\n)} \right)   ,
\end{align}
\end{subequations}
leads to the equation of motion
\begin{align}
 A m a_a =  &- (  A m \alpha ) \nabla_a \Phi + \hat{q} a_a f  +  \frac18 q^{tt}  a_a h_{tt}   + f \nabla_a \hat{q} + \hat{q} \nabla_a f + \frac18 h^{tt} \nabla_a q^{tt} + \frac18 q^{tt} \nabla_a h_{tt}  \\ \no 
&-  \frac14 q^{tt} a^{b} h_{ab} - \frac14 \dot{q}^{tt} h_{ta} - \frac14 q^{tt}\nabla_t h_{ta} - \frac14 q^{tt} a_a h_{tt}
\end{align}
in FNC. Inserting the results for the regular fields and their derivatives \eqref{eqs:h-reg} and making the substitutions, $q^{tt} = 4 m A$, $M := m(\Phi)  A(\Phi) $, and finally $ Q := \hat{q}(\Phi) $ we find that the local self-force is given by
  \begin{align}
  F_a^{\text{L}} &=   Q \nabla_a \Phi + 
  M^2 \left(-  \tfrac{11}{3}  \dot{a}_{a}   + \tfrac{1}{6}   
R_{at}  -2   \nabla_{ a}\Phi{} \nabla_t \Phi{} \right) + 
Q^2  \left( 
 \tfrac{1}{3} \dot{a}_{a}  + \tfrac{1}{6} R_{at}  \right) \nonumber \\
&  + Q M \left(11   a_{a}  \nabla_t \Phi{}   + 2   \nabla_{t} \nabla_{a}  \Phi{} \right)   +  \left(\a Q - M \a' \right)^2  \nabla_a \Phi \nabla_t \Phi.  %\left(  M^2 (\alpha(\Phi))^4  + 2   Q^2  \tfrac{\partial
%\alpha(\Phi)}{\partial \Phi}  +   M^2 \left(\tfrac{\partial \alpha(\Phi)}{\partial \Phi} \right)^2\right) \nabla_a \Phi \nabla_t \Phi.
\end{align}
Notice the similarity with the local self-force in a scalarvac spacetime. Notable though is the absence of the $Q^2 \nabla_a \Phi \nabla_t \Phi$ term. This term in the scalarvac calculation comes from  ${\sf m} \nabla_t h_{ta}$ through the term $\dot{q}^{tt} a_a$, where the scalarvac mass is defined as $q^{tt}_{\text{scalarvac}}= {\sf m}  := 4(m-q\Phi)$. Here, the analogous coupling in $q^{tt}\nabla_t h_{ta}$ has no charge squared term due to the difference in the definition of $q^{tt}$. % that extra minus four would give the value of 7 and reproduce exactly the EFT scalar result
Instead, we find a new coupling coming from the term $a_a \dot{Q} = a_a (\a Q - M \a')\dot\Phi $ in the derivative of the regular scalar field.
  
This expression can be expressed in  order-reduced  form by employing the background equation of motion $a_a = Q/M \nabla_a \Phi$ and its derivative $\dot a_a = \frac{Q}{M} \nabla_t \nabla_a \Phi -  \frac{\left( M \a' - \a Q \right)}{M} \nabla_t \Phi \nabla_a \Phi + 2 \frac{Q^2}{M^2} \nabla_t \Phi \nabla_a \Phi.$ Performing the order reduction and substituting the expression for the Ricci tensor in terms of the background scalar field $R_{ta} = \nabla_t \Phi \nabla_a \Phi + \orderof{s}$ gives 
 \begin{align}
 F_\a^{\text{L}} &=  Q w_\a^{\ \b} \nabla_\b \Phi  + \Bigg[ -\frac{11}{6} M^2 + \frac{23}{6}Q^2 + \frac13 \frac{Q^2}{M^2} \left(2 Q^2 - M ( M \a' - \a Q) \right) \no \\  &+  \frac{11}{3} M ( M \a' - \a Q ) +  (\a Q - M \a')^2 \Bigg] w_\a^{\ \b}  u^\c \nabla_\c \Phi  \nabla_\b \Phi  \no \\  &+ \left(\frac13 \frac{Q^3}{M} - \frac{5}{3} QM \right) w_\a^{\ \b}  u^\c \nabla_\c \nabla_\b \Phi
 \end{align}
 for the local self-force in covariant form.
 
  In addition to the local force acting at the current time, we also find a non-local tail part depending on the chronological past. 
The contribution from within the past light cone of the present is given by the tail force:
 \begin{align}
F_a^{\text{tail}} = &Q a_a f[\text{tail}]     + f[\text{tail}]  \nabla_a Q + Q \nabla_a f[\text{tail}]  + \frac12 M \a h_{tt}[\text{tail}]  \nabla_a \Phi + \frac12 M \nabla_a h_{tt}[\text{tail}]   \no \\ 
&-  M a^{b} h_{ab}[\text{tail}]  - M \a \dot\Phi h_{ta}[\text{tail}]  - M \nabla_t h_{ta}[\text{tail}]  - \frac12 M a_a h_{tt}[\text{tail}], 
\end{align}
which, after converting into covariant form and performing order-reduction, reads 
\begin{align}
F_\a^{\text{tail}}&  = w_{\a\b}  \Biggl[ 
-\a Q \nabla^\b \Phi f[\text{tail}] - \left( M \a'  - \a Q \right)\nabla^\b \Phi f[\text{tail}] + Q \nabla^\b f[\text{tail}] 
- Q \nabla^\c \Phi h^{\b}_{ \ \c}[\text{tail}]   \no \\ &- Q \nabla^\b \Phi h_{\c\d}[\text{tail}] u^{\c} u^{\d} + \frac12 M \left( \nabla^\b h_{\c\d}[\text{tail}]- 2 \nabla_\c h^{\b}_{\ \d}[\text{tail}]  \right)u^\c u^\d \Biggr].
\end{align}

The perturbed regular field also contributes to the evolution equation for the particle's mass
\begin{align}\label{eq:DM/dt}
\frac{D M}{d \tau} = & -Q u^\a \nabla_\a \Phi + 2 QM \left(\nabla_\c \nabla_\d \Phi - \frac12 F' g_{\c\d} \right) u^\c u^\d - \frac{1}{12} Q^2 R + Q^2 \left(\nabla_\a \Phi \nabla^\a \Phi + \frac12 F'' \right) \no \\ 
&+ (\a Q - \a' M)^2  (u^\a \nabla_\a\Phi)^2 - (\a Q - M \a') (u^\a \nabla_\a\Phi) f[\text{tail}]- Q u^\a \nabla_\a f[\text{tail}] 
\end{align}
as well as the evolution equation for the charge
{\small
\begin{align}\label{eq:DQ/dt}
\frac{D Q}{d \tau} = & \,\, Q' u^\a \nabla_\a{\Phi} - Q' Q''  (u^\a \nabla_\a{\Phi})^2   -  2 M Q'  \left(\nabla_\c \nabla_\d \Phi + \frac12 F' g_{\c\d} \right) u^\c u^\d -  Q Q' \left(\nabla_\a \Phi \nabla^\a \Phi  + \frac12 F'' \right) \no \\ 
	&+ \frac{1}{12} Q Q' R + Q'' u^\a \nabla_\a \Phi f[\text{tail}] + Q'   u^\a \nabla_\a  f[\text{tail}],
\end{align}}
where $Q' = \a Q - M \a'$ and $Q'' = 2 \a' Q + \a \left( \a Q - M \a' \right) - M \a''$. Starting with initial conditions $Q(\Phi(0,\vec{x}))=0$, Eq.~\eqref{eq:DQ/dt} allows for a non-zero charge at some later time. In other words, the background scalar field can \emph{induce} a charge on the world line dynamically.  This suggests that scalarization phenomena are present in the EMRI problem as well. 

%%%%%%%%%%%%%%%%%%%%%%%%%%%%%%%%%%%%%%%%%%%%%%%%%%%%%%%
\section{Scalar-tensor self-force in a stationary black hole background }\label{sec:bh}
\subsection{Field equations}\label{sec:bh-field-eqns}
%%%%%%%%%%%%%%%%%%%%%%%%%%%%%%%%%%%%%%%%%%%%%%%%%%%%%%%
As mentioned in the introduction,  the Kerr geometry accompanied by an external, constant scalar field represents the most general stationary, axisymmetric and vacuum solution to generic scalar-tensor theories \cite{Sotiriou:2011dz}.  Non-stationary solutions for the scalar field configuration such as those arising from scalar-tensor theories with cosmological evolution or those with potentials leading to non-uniform asymptotic scalar field configurations may create hairy configurations. Though possible, their existence is highly constrained by measurements of period decays in binary systems displaying the absence of dipole radiation as predicted by hairy configurations \cite{Damour:1998jk}.  Additionally, asymptotic flatness requires that the scalar potential and its first derivative with respect to the background scalar field vanish far from the black hole. To be consistent with a stationary black hole configuration, we also demand that the background scalar field be fixed at the constant value $\Phi_0$ .  Consequently, the background Ricci tensor vanishes, which implies that the stress energy of the scalar field is also zero. The background Einstein field equations are therefore given by the vacuum field equation $R_{\mn} = 0$ giving rise to the Kerr family of black hole solutions for stationary, axisymmetric backgrounds.  

As before, we consider a non-spinning massive point-like object moving in the black-hole spacetime of scalar-tensor theory. The particle's stress-energy generates perturbations $\gamma_{\a\b}$ of the background geometry.   The Lorenz gauge $\nabla_\a \gamma^{\a\b}=0$ field equations governing perturbations of the black hole with a constant scalar field background take the completely decoupled form
\begin{subequations}\label{eq:bh-eqns}\begin{align}
\Box \gamma_{\a\b} + 2 R_{\a \ \b}^{\ \m \ \n}\gamma_{\mn} &= - 16 \pi M \int \delta_4(x,z) u_\a u_\b d \tau, \\
\Box f - \mu^2 f &= - 8 \pi Q \int  \delta_4(x,z) d \tau ,
\end{align}
\end{subequations}
 where $M:=m(\Phi_0) A(\Phi_0)$  and $Q:=-\a(\Phi_0) M$ are the constant mass and charge of the point-like object, respectively, and $\mu^2 = F''(\Phi_0)$ is the mass of the scalar field.  The equation governing the metric perturbations is the well-known Lorenz gauge wave equation and the scalar field equation is the curved spacetime massive wave equation. 
 
 The retarded solutions to  the above decoupled system of equations  are given in terms of the diagonal Green functions by
 \begin{subequations}
 \begin{align}
 \gamma_{\a\b}(x) &= 4 M \int G_{\a\b\c\d}(x,z(\tau)) u^\c u^\d \, d \tau , \\ \no
 f(x) &= Q\int G(x,z(\tau)) \, d \tau.  
 \end{align}
 \end{subequations}
 Each solution is then separately regularized according to its own DW singular field to obtain the regular fields responsible for the self-force.  The expressions for the singular/regular decoupled fields are widely available in the literature \cite{PPV}.
 
 \subsection{Equation of motion}
For a stationary black hole background spacetime with a constant scalar field $\nabla \Phi =0$ the leading order motion of the particle is geodesic in the background geometry $a^\a = 0$.   The next-to-leading order motion is given by the first order self-force which is simply the sum of the vacuum gravitational and scalar tail self-forces
\begin{align}
F^\a_{\text{self}} &= F^\a_{\text{gravity}}[\text{tail}] + F^\a_{\text{scalar}}[\text{tail}] \no \\
&=  \frac12 M  w^\a_{\ \b}  \left( \nabla^\b h_{\c\d}[\text{tail}]- 2 \nabla_\c h^{\b}_{\ \d}[\text{tail}]  \right)u^\c u^{\d} + Q w^\a_{ \ \b} \nabla^\b f[\text{tail}].
 \end{align}
The lack of local terms in the equation of motion is due to the geodesic leading-order motion of the particle and the  absence of coupling terms in the linearized field equations.   
The mass evolution is dictated by the equation
\begin{equation}\label{eq:BH-dMdt}
\frac{D M}{d\tau} = \frac12 Q^2 \mu^2 -Q u^\a \nabla_\a f[\text{tail}],
\end{equation}
and the charge is found to evolve according to the equation
\begin{equation}\label{eq:BH-dQdt}
\frac{D Q}{d\tau} = -\frac12 Q Q' \mu^2   + Q'  u^\a \nabla_\a f[\text{tail}].
\end{equation}
Interestingly, we find that the parameters characterizing particle experience a local self-field correction, which depends on the mass of the scalar field $\mu$, the charge $Q=-M'(\Phi_0) $, and the profile of the charge  $Q' = -M''(\Phi_0)$ at the current time, as well as a history dependent correction coming from the scalar field sourced by the particle in its past light cone. However, for realistic cosmological values, which can be as low as the Hubble scale $\mu \sim 10^{-33} \text{eV}$, the local change in the mass and charge of the particle will be suppressed relative to the tail piece.  
 
\section{Conclusion and future directions} 
 In this paper we derived the regular field and the equation of motion for a massive point-particle in the Einstein frame of scalar-tensor theory to first-order in the mass-ratio. We find that the self-forced evolution in scalar tensor theory is different from the evolution of a massive, scalar charge in an scalarvac spacetime of general relativity. For non-stationary scalar field backgrounds, the leading-order motion is accelerated in the background spacetime and the first-order self-force correction takes the form of a set of local and non-local terms.  The non-local, ``tail'', component of the force is characterized by non-diagonal Green functions which result from the coupling in the field equations. The local ST self-force resembles the local self-force on a charged, massive particle in a scalarvac spacetime with additional couplings due to the time variability of the particle's charge in addition to its time variable mass.   The reason is that in ST theory, the strong equivalence principle is violated and the motion is sensitive to the internal constitution of the body. Here we used the point-particle approximation, where the details of the body's internal structure are coarse grained out.  However, the mass and charge of the object carry a scalar field dependence, and this creates fundamentally different world line dynamics. Furthermore,  the evolution of the mass and charge of the particle is influenced both locally and non-locally by the regular self-field. This is true even for black holes with stationary scalar backgrounds, where the mass and charge of the particle evolve locally if the scalar field has mass and non-locally through the regular part of the scalar perturbation.  This has consequences for the floating orbit scenario. It is evident from our equations that self-field corrections to the particle's mass may cause a violation of the super-radiance condition or lead to an imbalance in the radiation fluxes, precluding any prolonged dynamical floating of the orbit.  Cardoso et al. \cite{Cardoso:2011xi} estimated the effect of ingoing negative scalar radiation flux at $\dot{J^-}(\mathscr{J}^+)$ on the mass of the large black hole, but they neglected to mention the possibility for the scalar particle to possess a dynamical mass. To estimate whether the particle's mass variation is important we can compare the timescale on which the particle's mass varies to the timescale on which the black hole's mass secularly decreases. When the orbit is in a floating configuration, the negative scalar flux at the event horizon is balanced by the flux of outgoing gravity waves at $\mathscr{J}^+$. This cannot last forever due to the decrease in the black hole's mass 
as the scalar field extracts energy from it. The timescale for the extraction is roughly $T_{\rm bh} \sim M_{\rm bh}^3/M^2$, where $M_{\rm bh}$ is the mass of the large Kerr black hole and 
$M$ is the mass of the particle. The timescale governing mass changes to the small body due to the self-force is roughly $T_{\rm sb} =  M/(Q^2 \mu^2) \sim 1/(M\mu)^2$ when ignoring the tail contribution. For self-force effects to be important they must occur before or at least around the time that significant mass loss of the black hole from the negative flux occurs, or  $T_{\rm sb} \lesssim T_{\rm bh}$ or $M/M_{\rm bh} \lesssim M_{\rm bh}^2 \mu^2$. Thus, self-force effects become important for a large region of the parameter space. 
 
 The self-force for ST  black holes with trivial scalar configurations is also different than in GR. In vacuum GR, the force is completely non-local and involves the tail force due to the metric perturbation only. In scalar-tensor theory, even with a constant background scalar field solution, the scalar perturbation will also influence the motion: a point particle orbiting an ST black hole experiences both the metric and scalar tail forces.   These deviations from the GR self-force will likely have observational consequences. The self-force is known to cause a shift in the innermost stable circular orbit, alter the rate of periastron advance for eccentric orbits \cite{Barack:2011ed}, and influence the renormalized redshift of photons sent from the particle to distant observers \cite{DiazRivera:2004ik}. 
 
  Next steps include computing observational effects in ST theory and using the results to constrain its parameters.  For instance, one can imagine setting up a numerical calculation of the self-force on a point-particle around a star or black hole in ST theory using the mode-sum method.  This will require formulating the coupled field equations in a fashion where the modes of the retarded field can be numerically determined and the mode representation for the singular field can be subtracted. Linz and collaborators have shown that the derivatives of the singular fields for a scalarvac spacetime completely decouple and that the regularization parameters for the force are the sum of the decoupled parameters \cite{Linz:2014vja}. It is reasonable to expect that the singular field gradients will also be decoupled in the present case, as the equations have the same form, and this will certainly alleviate some of the computational challenge as only the retarded field will need to be solved from coupled equations. One could also imagine solving the decoupled field equations \eqref{eq:bh-eqns} to compute the self-force and mass/charge evolution for a particle outside a Kerr black hole in ST theory. One could then choose initial data and scalar field parameters corresponding to a floating orbit of Ref.~\cite{Cardoso:2011xi} and compute the self-force and mass evolution  to assess the stability of the orbit. 
 
\begin{acknowledgments} 
We thank Eric Poisson, Chad Galley, Sam Gralla, Luis Lehner, and Alexandros Gezerlis  for helpful comments and discussions.  
\end{acknowledgments}

 \end{document}